\documentclass[%
    aip,
     amsmath,amssymb,floatfix,
     preprint,%
     reprint,%
]{revtex4-1}

\usepackage{graphicx}
\usepackage{bm}
\usepackage{siunitx}
\usepackage{xcolor}
\usepackage{float}
\makeatletter
\let\newfloat\newfloat@ltx
\makeatother
\usepackage{algorithm}
\usepackage{algpseudocode}
\usepackage{amsmath}

\usepackage[T1]{fontenc}
\usepackage{mathptmx}
\usepackage[hyperfootnotes=false]{hyperref}
\usepackage{amsmath}
\DeclareMathOperator*{\argmin}{arg\,min}
\usepackage{booktabs}
\usepackage{makecell}
\usepackage{natbib}

\newcommand{\appautoref}[1]{{\renewcommand{\sectionautorefname}{Appendix}\autoref{#1}}}

\begin{document}

\title[Robust Control of ECH Deposition Profiles on DIII-D]{Robust Control of ECH Deposition Profiles on DIII-D}

\author{H.J. Farre-Kaga}\email{hf8585@princeton.edu}
\affiliation{The two authors contributed equally to this paper}
\affiliation{Princeton University, Princeton, NJ, USA}
\affiliation{Princeton Plasma Physics Laboratory, Princeton, NJ, USA}
\author{A. Rothstein}
\affiliation{The two authors contributed equally to this paper}
\affiliation{Princeton University, Princeton, NJ, USA}
\author{K. Yasoda}
\affiliation{Princeton University, Princeton, NJ, USA}
\author{J. Lestz}
\affiliation{University of California Irvine, Irvine, CA, USA}
\author{N. Chen}
\affiliation{Princeton University, Princeton, NJ, USA}
\author{S.K. Kim}
\affiliation{Princeton Plasma Physics Laboratory, Princeton, NJ, USA}
\author{A. Jalalvand}
\affiliation{Princeton University, Princeton, NJ, USA}
\author{E. Kolemen}\email{ekolemen@pppl.gov}
\affiliation{Princeton University, Princeton, NJ, USA}
\affiliation{Princeton Plasma Physics Laboratory, Princeton, NJ, USA}

\begin{abstract}

Electron Cyclotron Heating (ECH) is a key actuator in DIII-D and future tokamaks that provides auxiliary heating, localized current drive for scenario development and MHD stability, and even impurity pump-out. Due to its control flexibility and applications, a gyrotron optimization algorithm was developed to multitask and fine-tune the deposition location and heating power of each gyrotron while providing robustness to hardware failure. The ECH Optimization (ECHO) algorithm finds the optimal gyrotron mirror angle and power to achieve a target ECH radial deposition profile. This optimization is accomplished in real-time using a parallelized neural network surrogate of the \texttt{TORBEAM} code combined with a genetic optimizer. The algorithm has been deployed in a DIII-D experiment and has been validated with experimental ECE measurements and post-experiment offline ray tracing, showing the algorithm’s reliability despite gyrotron failures and significant changes to plasma parameters. 

\end{abstract}
\maketitle

\section{Introduction}\label{sec:intro}

Tokamaks are one of the most promising paths towards realizing commercial fusion power plants (FPP)\cite{han_sustained_2022}, but they require external heating sources to efficiently achieve energy positive conditions\cite{poli_external_2014,rodriguez-fernandez_overview_2022} and external current drive to sustain steady state operation\cite{luce_role_2011}. Electron cyclotron heating (ECH) is a localized heating and current drive source and is planned as the primary heating source for ITER\cite{omori_overview_2011,harvey_electron_1997}. This is because ECH is effective in lower toroidal field tokamaks that operate at densities below the cutoff. ECH has also been tested and validated on many tokamaks, proving the technology is ready to be deployed in a commercial FPP environment\cite{cengher_status_2020,joung_kstar_2024,verhoeven_design_2003,leuterer_ecrh_2001,xu_development_2016}. ECH is also applied to stellarators\cite{turkin_current_2006}, another highly promising fusion reactor concept.

Radio frequency (RF) heating sources are desirable for FPPs due to poor scaling of neutral beam injection (NBI) to reactor relevant regimes\cite{hopf_neutral_2021} and their large port space requirements. While negative-ion-based NBI (N-NBI) systems can overcome some of these physical limitations, N-NBI has not seen the same level of rigorous technological demonstration as RF heating, having been deployed on just two machines\cite{oka_operation_2001,kuriyama_operation_1998}. Other RF heating sources like helicon\cite{praterApplicationVeryHigh2014,pinskerFirstHighpowerHelicon2024a}, ion cyclotron resonance heating (ICRH)\cite{casson_predictive_2020,monakhov_assessment_2025} and lower hybrid current drive (LHCD)\cite{tuccillo_progress_2005,leppink_high-field_2025,rutherfordPredictions212026} have a number of benefits and have been tested on fusion reactors, but all of these suffer from the same limitation of lower frequency ranges making mirror-like launchers impossible. Instead, they use antennae to release the RF waves into the vessel\cite{chappuis_design_2005,kaye_progress_1993,wukitch_high_2017,kim_high_2019}. Without mirror-like launchers, the ability to change the deposition location can only be adjusted by changing plasma parameters\cite{choi_simulation_2024} like $B_T$, the RF frequency, or ion minority species being heated. These are primary engineering parameters for a tokamak that cannot be adjusted in steady-state operation, and are therefore not suitable for real-time heating deposition control. This limits their usage for critical instability-specific control\cite{ramponi_iter_2007} and scenario fine-tuning. 

The shorter wavelength of ECH allows it to be injected into the plasma through a steerable mirror, directly changing the ray trajectory and eventual deposition location\cite{prater_design_1997}. The ECH waves are generated by gyrotrons, which currently cannot penetrate into high toroidal field plasmas, although the technology is rapidly improving and could soon bridge this gap\cite{ikeda_multi-frequency_2023}. However for now, ECH is prime for lower toroidal field machines such as ITER with $B_T=$\SI{5.3}{T}\cite{huguet_magnet_1997}. While high densities can also stop electron cyclotron waves from penetrating the plasma\cite{wesson_tokamaks_2011} in X-mode polarization, O-mode waves can be absorbed by higher density plasmas at the expense of a lower absorption efficiency\cite{minashin_model_2015}. 

ECH has been demonstrated in extensive applications for tokamaks: steady state current drive\cite{luce_role_2011,ushigusa_noninductive_2002,polevoi_assessment_2008,turkin_current_2006}, NTM control\cite{sauter_requirements_2010,zohm_control_2007,kolemen_real-time_2013, farre-kagaInterpretingAIFusion2026}, ECH-assisted plasma breakdown\cite{granucci_plasma_2011,jackson_plasma_2010,liu_ecw_2024}, current profile control\cite{pajares_integrated_2019}, density pump-out\cite{weisen_particle_2001,idei_experimental_1995,wang_understanding_2017}, impurity control\cite{tamura_observation_2017,sertoli_interplay_2015,puiatti_simulation_2003}, sawtooth control\cite{graves_recent_2011,maraschek_active_2005,felici_integrated_2012}, control of Alfv\'en Eigenmodes \cite{sharapovEffectsElectronCyclotron2018,vanzeelandReversedShearAlfven2008a}, and improving robustness of RMP ELM suppression\cite{hu_effects_2024,logan_access_2024}. However, limited research has been devoted to balancing these competing control tasks that all use a limited amount of total EC power, such as ECH multitasking\cite{pajares_integrated_2019}. Work has been done to develop ITER's control system\cite{vu_progress_2026} to achieve multiple control tasks, but does so in a framework that treats all gyrotrons as independent instead of a collective heating units that has a total heating and current drive deposition profile.

\begin{figure}
    \centering
    \includegraphics[width=0.49\textwidth]{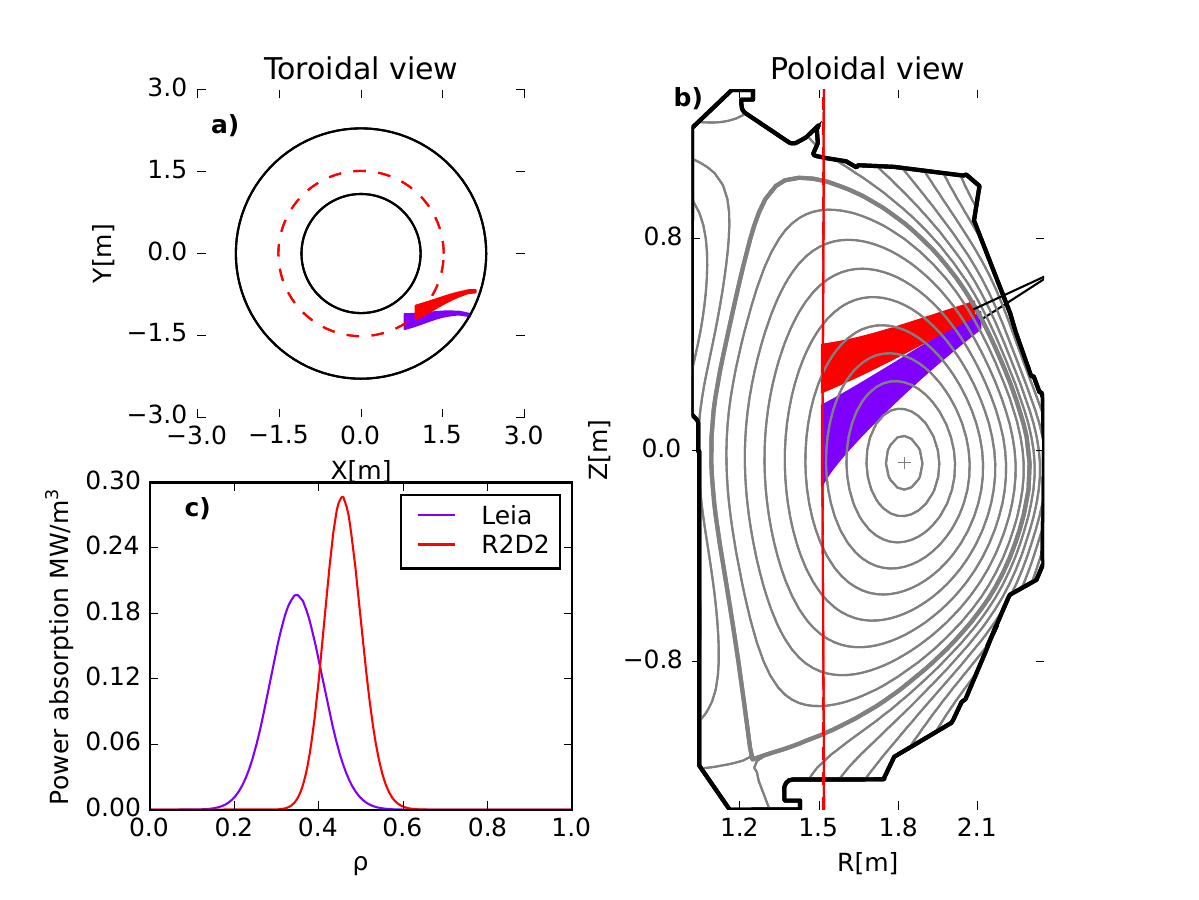}
    \caption{Example ECH ray trace for a given DIII-D equilibrium for the Leia gyrotron (purple) and the R2D2 gyrotron (red), both in X-mode polarization. \textbf{a)} Toroidal cross section with the resonant magnetic field location (dashed red line) and two ECH ray trajectories (red and purple shaded regions) with different toroidal angle launch locations. \textbf{b)} Poloidal cross section of ECH ray trace with resonant absorption location (vertical red line) and ray trajectories with different poloidal angles (red and purple shaded regions). \textbf{c)} ECH deposition profile for the two given gyrotrons across spatial toroidal flux coordinate $\rho$.}
    \label{fig:ray_trace}
\end{figure}

A basic visualization of an ECH ray trace is shown in \autoref{fig:ray_trace}, plotting two gyrotron beams with different mirror locations, toroidal injection angles, and poloidal injection angles. In this paper, $\rho$ is defined as the square root of the normalized toroidal magnetic flux. The exact trajectory of the waves depends on the magnetic equilibrium and plasma profiles, and is calculated from the \texttt{TORBEAM}\cite{poli_torbeam_2018} ray trace code. As the toroidal and poloidal mirror angles change, the trajectory of ECH and the eventual absorption location will change. At larger electron densities and larger poloidal injection angles, the amount the ECH rays bend and are refracted increases until the ECH density cutoff limit is reached when the rays are perfectly reflected. At large toroidal injection angles, the ECH heating also produces electron cyclotron current drive (ECCD), which can be in the direction of the plasma current or against it. Finally, the total power delivered can be changed during the shot by modulating the individual gyrotrons, turning them on and off rapidly, to emulate a gyrotron delivering a fraction of the total power. For example, turning an individual gyrotron with \SI{1}{MW} off for \SI{2}{ms} and then on for \SI{2}{ms} approximates a similar gyrotron continuously on with \SI{0.5}{MW}. 

Significant previous work has analyzed the aiming requirements\cite{cirant_crucial_2005}, especially in the case of tearing mode suppression\cite{farina_eccd_nodate}, and a common finding is that real-time ECH ray tracing will be required to achieve desired ECH-related goals. For instance, suppressing instabilities could require depositing the ECH at a surface of constant rho or q, whose absolute spatial location may evolve in time as the equilibrium evolves. To this end, the \texttt{TORBEAM}  code has a reduced version that is real-time capable\cite{poli_torbeam_2018}, running in approximately \SI{10}{ms}, with exact timing depending on implementation and hardware details. While this is effective for basic aiming, the real-time \texttt{TORBEAM} code only solves for the location of maximum absorption and does not calculate the full ECH and ECCD deposition profiles. In addition, the time scale of \SI{10}{ms} is not fast enough to rapidly calculate the optimal mirror angle corresponding to a desired deposition location, which needs multiple real-time \texttt{TORBEAM} runs. There is an accelerated version of \texttt{TORBEAM} that calculates full deposition profiles, but that runs in approximately \SI{75}{ms}. 

To address the remaining gaps in real-time ECH ray tracing, a machine learning (ML) surrogate model of the full-fidelity ray trace code has been trained, achieving computation times a fraction of the real-time ray trace codes. An ML surrogate model\cite{irvin_surrogate_2025} of the TORAY code\cite{prater_benchmarking_2008} was developed to provide quicker ray trace results. However, this model was developed for reactor design workflows, not real-time application. As such, the uncertainties of this model were too large for real-time application as it was trained to predict for any ECH launcher location with large parameter space of plasma conditions. While surrogates trained on a large parameter space may be better for extrapolation, machine-specific surrogates can be made more accurate and to run faster, which is the key requirement for real-time control applications. Therefore, an ML surrogate model for \texttt{TORBEAM} was developed for the KSTAR ECH launcher geometry and plasma conditions\cite{rothstein_torbeamnn_2025}, focusing on the maximum absorption location for real-time steering and thus achieving more accurate model predictions. Due to lower compute power available in the KSTAR PCS compared to DIII-D, this surrogate model only calculates the peak absorption location, not the full ECH deposition profile. An upgraded surrogate model for \texttt{TORBEAM} that calculates the full deposition profile in DIII-D is presented in this paper.

This work develops a framework for coordinated, advanced ECH control to rapidly and accurately achieve a desired total ECH deposition profile. Instead of deciding on individual gyrotron tasks as in \citet{vu_progress_2026}, this framework intelligently coordinates all available gyrotrons for a single task of precisely matching the target deposition profile. To do this, a new ML surrogate model of the \texttt{TORBEAM} code was developed for the DIII-D launcher geometry, producing full ECH deposition profiles of each individual gyrotron. This surrogate model runs fast enough to compute a look-up table of possible injection angles, which is then searched by a genetic algorithm-based optimizer to find the optimal gyrotron mirror angles and powers. These optimized angles and powers best match the target total ECH deposition profile given the hardware constraints and plasma conditions. DIII-D gyrotrons cannot directly adjust their output power in real time, but they can be modulated at different duty cycles (the ratio between the time spent on vs off), which effectively produces a real-time adjustable gyrotron power. 

In \autoref{sec:ECH-problem} we describe the ECH control problem with some simplifying assumptions to make the problem more tractable in real-time, followed by a detailed description of the TorbeamNN surrogate model and the genetic algorithm-based optimizer to run real-time in the PCS. Next in \autoref{sec:applications} are the control results from DIII-D experiments, with applications to basic ECH profile control, demonstrations in changing plasma conditions, and fault handling. Finally, we have a summary and concluding discussion in \autoref{sec:conclusion}. 

\section{ECH Control Problem and Real-Time Models}\label{sec:ECH-problem}

\begin{figure*}
    \centering
    \includegraphics[width=\linewidth]{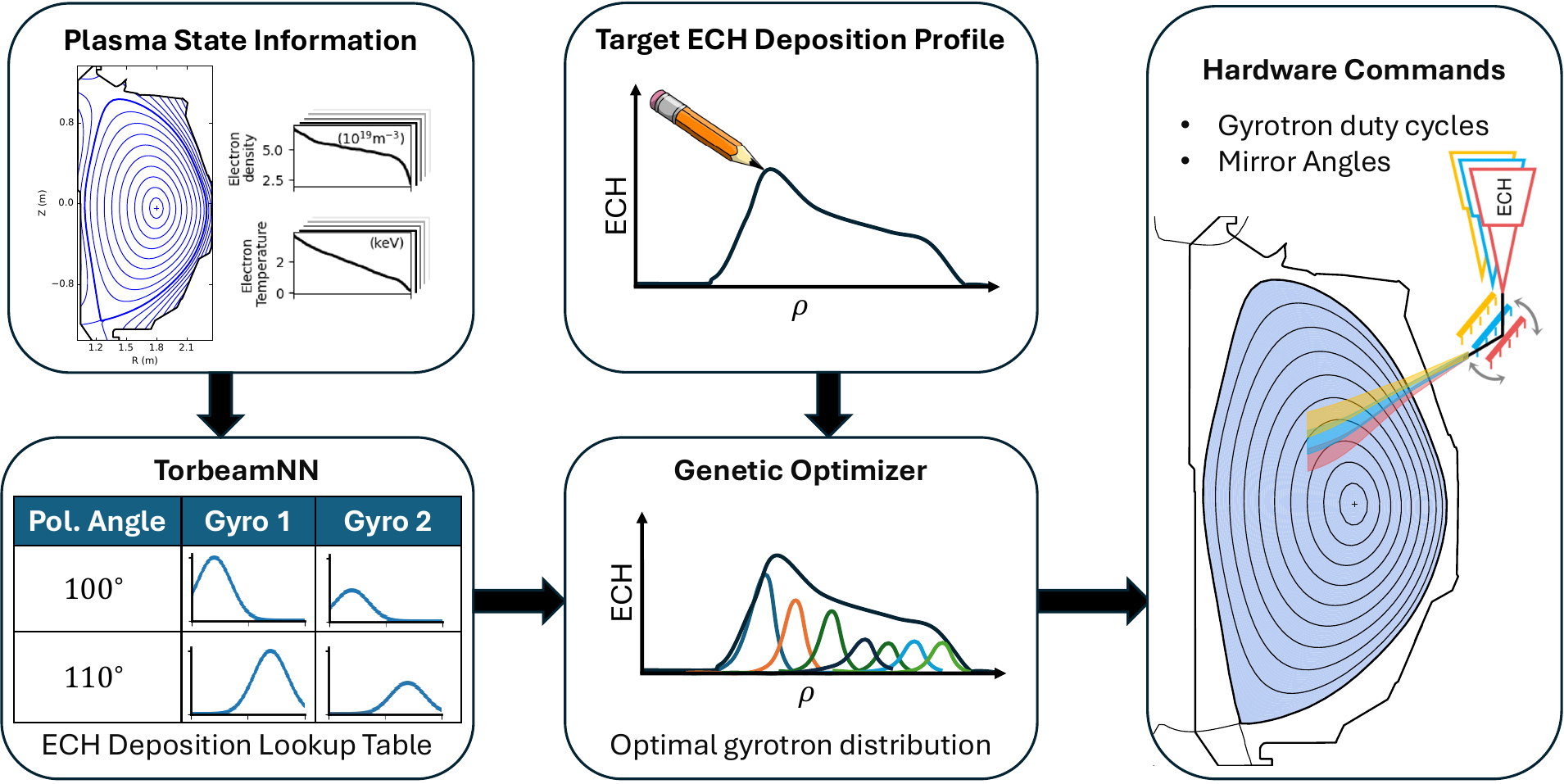}
    \caption{Overview of the ECHO algorithm. The \textbf{plasma state information} including profiles and equilibria are first measured. This is passed to \textbf{TorbeamNN}, which is run in parallel for each gyrotron and several poloidal angles to make a lookup table of individual deposition profiles. The \textbf{genetic optimizer} then uses a user-defined \textbf{target ECH deposition profile} and the lookup table to find the optimal gyrotron configuration. These duty cycles and mirror angles are then sent as \textbf{hardware commands} to actuate on the plasma. The entire sequence including generating the TorbeamNN lookup table must be run every real-time cycle to account for changing plasma conditions.}
    \label{fig:echo-overview}
\end{figure*}

The goal of the ECH optimizer (ECHO) is to deliver an ECH deposition profile that most closely matches a target as visualized in \autoref{fig:echo-overview}. ECHO has knowledge of available power from each gyrotron (including mid-shot hardware failures), hardware and user-defined constraints such as gyrotrons that can't modulate, and real-time plasma state parameters. The algorithm uses this information to find the optimal ECH power and ECH deposition location. First, the plasma state is measured and used by TorbeamNN to create a lookup table for each gyrotron, mapping ECH mirror angle to ECH deposition profile, described in full in \autoref{subsec:torbeamNN}. Next, the genetic optimizer uses this lookup table to find the duty cycles and angles for each gyrotron to achieve a total ECH deposition profile that best matches the target. A full explanation of the genetic optimizer is given in \autoref{subsec:genetic_algo}. Finally, the optimized duty cycles and angles are passed to the hardware to adjust the ECH injected into DIII-D to achieve the target ECH deposition profile. All the calculations in \autoref{fig:echo-overview} must be run every real-time cycle to accurately model changing plasma conditions such as profile changes. 

This section begins with a description of the ML surrogate model TorbeamNN in \autoref{subsec:torbeamNN}, followed by an explanation of the genetic optimizer in \autoref{subsec:genetic_algo}. Details on the plasma control system (PCS) implementation are given in \autoref{subsec:PCS}, and finally some assumptions and simplifications of the ECH control profile are discussed in \autoref{subsec:assumptions}.

\subsection{TorbeamNN Lookup Table}\label{subsec:torbeamNN}

Modeled off of the \texttt{TORBEAM} code\cite{poli_torbeam_2018} and a similar model developed for KSTAR applications\cite{rothstein_torbeamnn_2025}, the DIII-D model of TorbeamNN takes plasma state parameters as input and produces ECH deposition profiles with the full list of inputs and outputs listed in \autoref{tab:torbeamNN}. Note that it is possible to reduce the full extent of input parameters needed, however some estimate of the equilibrium shape, electron temperature, and electron density profiles are required for ECH ray tracing. We parameterize each ray trace to produce a Gaussian deposition profile. This is a generally reasonable assumption, but can break down near the plasma core and edge where skewness in the deposition profiles can become significant. Additionally, predicting the three parameterized outputs instead of a full radial profile requires far less computation and is ideal for real-time applications. We also perform a PCA reduction of the $T_e$ and $n_e$ profiles to four principal components to further reduce the model's computation. This has a reconstruction $R^2$ value of 0.9991, so little information is lost despite a significant input reduction. The poloidal angle is scanned by ECHO, and the toroidal angle comes from the ECH mirror hardware. This is because on DIII-D, the poloidal angle can be adjusted in real-time, while the toroidal angle is fixed and can only be adjusted between shots. The poloidal angle can change at an approximate rate of $5^\circ$ per \SI{100}{ms}, which corresponds roughly to a $\rho$ change of $0.075$ per \SI{100}{ms}. All shape and equilibrium parameters come from RT-EFIT\cite{ferron_real_1998}, a real-time code that uses magnetic diagnostics to reconstruct the plasma equilibrium. The $T_e$ and $n_e$ profiles come from RTCAKENN\cite{shousha_machine_2023}, an ML surrogate model of the automated kinetic equilibria reconstruction tool CAKE that uses magnetic diagnostics, Thomson Scattering, CER, and MSE to reconstruct plasma equilibrium profiles. 

To create the lookup table described in \autoref{fig:echo-overview}, TorbeamNN is run in parallel for each gyrotron and several poloidal angles. The angle range and steps are selected by the user, typically scanning a full range in $\rho$. Even though (at the time of writing) each gyrotron on DIII-D has the same frequency and $(R,Z)$ launch geometry, TorbeamNN is run separately for each gyrotron because they have independent toroidal launch angles which are typically different for every gyrotron. 

\begin{table}
    \centering
    \begin{tabular}{|c|c|}
        \hline\textbf{Input} & \textbf{Description} \\\hline
        $\theta$ & Poloidal ECH mirror angle  \\ 
        $\phi$ & Toroidal ECH mirror angle\\
        $R$ & Major radius \\
        $a$ & Minor radius \\
        $Z$ & Vertical position \\
        $I_P$ & Plasma current \\
        $B_T$ & Toroidal field \\
        Gapbot & Distance from plasma to lower vessel wall \\
        Gapin & Distance from plasma to inner vessel wall \\
        Gapout & Distance from plasma to outer vessel wall \\
        Gaptop & Distance from plasma to upper vessel wall\\
        $\kappa$ & Plasma Elongation \\
        $l_i$ & Plasma inductance \\
        $\delta_{bot}$ & Lower triangularity \\
        $\delta_{top}$ & Upper triangularity \\
        $V$ & Plasma Volume \\
        $T_e$ & Electron temperature profile \\
        $n_e$ & Electron density profile \\
        \hline\hline\textbf{Output} & \textbf{Description} \\\hline
        $\mu_{ECH}$ & Center of ECH Gaussian (in $\rho$) \\
        $\sigma_{ECH}$ & Standard deviation of ECH Gaussian (in $\rho$)\\ 
        $A_{ECH}$ & Peak height of ECH Gaussian (in \si{MW/m^3})\\\hline
    \end{tabular}
    \caption{Inputs and outputs of DIII-D TorbeamNN model used as part of ECHO. }
    \label{tab:torbeamNN}
\end{table}

The training dataset consists of 2,234,088 \texttt{TORBEAM} ray trace calculations from 4,168 unique plasma shots, which are split into training, testing, and validation data with a $70:15:15$ split. The split is done on a per shot basis to avoid potentially leaking information from training to validation sets. After hyperparameter tuning, the final trained model uses 4 dense layers with 96 units each using a ReLU activation function, where each dense layer is followed by a batch normalization layer and then a dropout layer with rate $0.1$. The final layer is a simple dense layer with a linear activation function. We use a batch size of 3716, an L2 regularization of $1.23\times 10^{-4}$, initial learning rate of $1.72\times 10^{-3}$ and the Adam optimizer. We train for up to $500$ epochs, but stop after $20$ epochs without validation loss improvement, which typically occurred around epoch $50$. 

\begin{table*}
    \centering
    \begin{tabular}{l ccc ccc} 
        \toprule
         & \multicolumn{3}{c}{\textbf{X-mode}} & \multicolumn{3}{c}{\textbf{O-mode}} \\
        \cmidrule(lr){2-4} \cmidrule(lr){5-7} 
        \textbf{Parameter} & Training & Validation & Testing & Training & Validation & Testing \\
        \midrule
        $\mu_{ECH}$  & 0.96 & 0.95 & 0.95 & 0.99 & 0.98 & 0.98 \\
        $\sigma_{ECH}$  & 0.74 & 0.70 & 0.69 & 0.91 & 0.89 & 0.89 \\
        $A_{ECH}$ & 0.82 & 0.78 & 0.75 & 0.90 & 0.87 & 0.83 \\
        \bottomrule
    \end{tabular}
    \caption{$R^2$ correlation values across both models and training, validation, and test datasets. }
    \label{tab:r2-values}
\end{table*}

\begin{figure*}
    \centering
    \includegraphics[width=\linewidth]{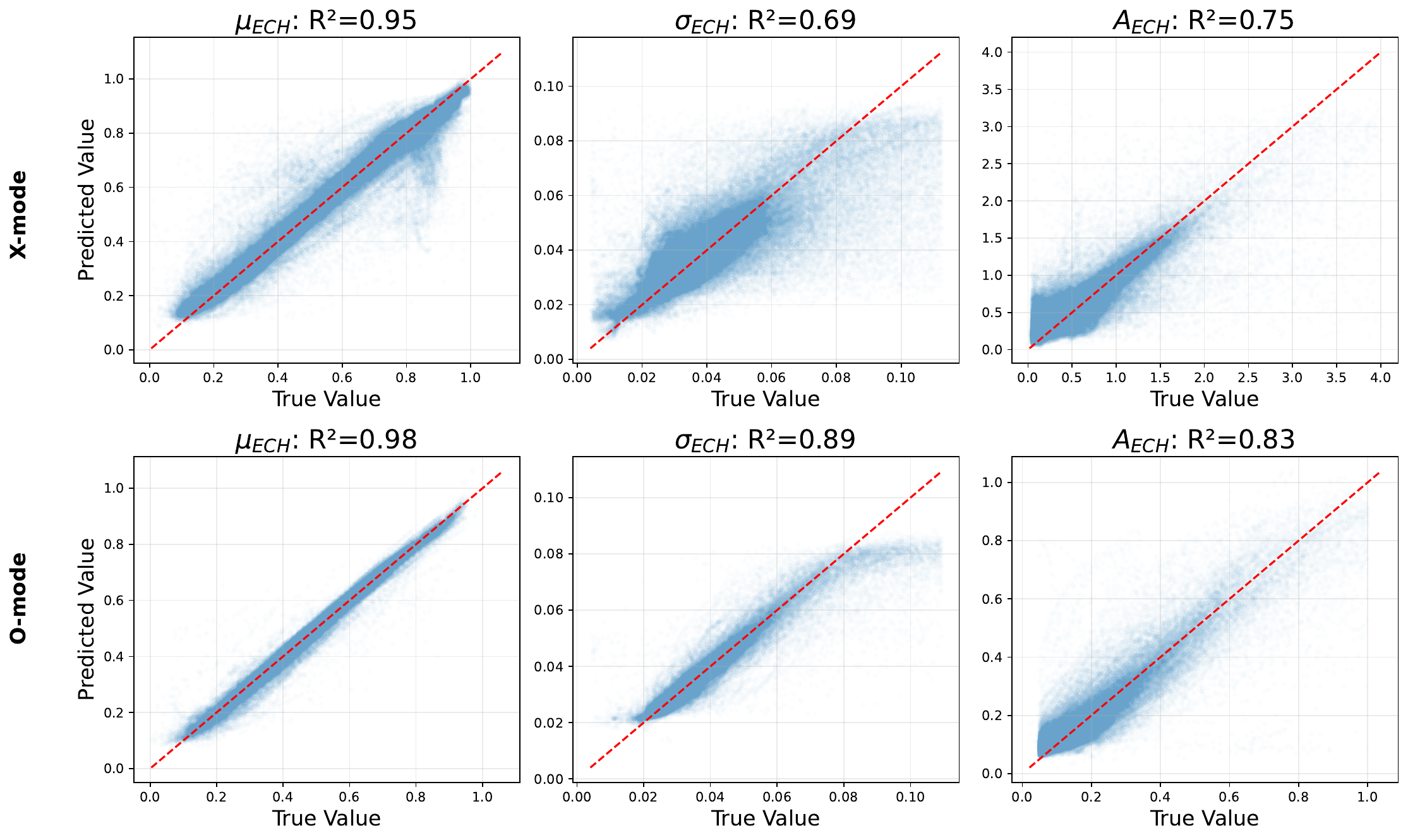}
    \caption{True (horizontal axis) versus predicted values (vertical axis) of $\mu_{ECH}$, $\sigma_{ECH}$, and $A_{ECH}$ for the X-mode (first row) and O-mode (bottom row) TorbeamNN models. Red dashed lines represent a perfectly correct model.}
    \label{fig:r2-scatter}
\end{figure*}

The training, validation, and testing $R^2$ values are provided in \autoref{tab:r2-values} and scatterplots of the test dataset comparison are in \autoref{fig:r2-scatter}. The drop offs from Training to Validation and Validation to Testing are relatively minor, so we find the model is accurate with never-before seen data, which is required for real-time use. We also see that O-mode absorption is generally much more accurate than X-mode absorption, which makes physical sense as DIII-D does not achieve high enough densities to reflect O-mode ECH. As O-mode does not refract significantly on DIII-D, the angle of incidence of the rays when they reach the resonance location is nearly entirely determined by the poloidal launch angle which makes the ray tracing problem inherently simpler to learn. We also find that both models are more accurate at predicting the central absorption location $\mu_{ECH}$ than either $\sigma_{ECH}$ or $A_{ECH}$. This suggests that physical ray trace is easier to learn than the absorption efficiency for a given plasma state, which is in line with the complexity of the physics of ray tracing compared to power absorption. Calculating heating and current drive requires accounting for kinetic effects that are physically more challenging to capture. For example, $\mu$ is already highly correlated with poloidal mirror angle and is much easier to learn. A possible explanation for poor performance of $\sigma$ at small and large values is that Gaussians may become skewed near the axis and near the edge. These imperfect Gaussians will lead to poor quality training data and consequently be much harder for the ML surrogate to learn. 

Looking at the scatterplots of predicted values in the test set in \autoref{fig:r2-scatter}, one interesting feature is the under-prediction of wide Gaussians where the real $\sigma_{ECH}$ is larger than the predicted value (scatterplot points fall below the red dashed line). This makes sense as large widths are poorly represented in the training dataset as this condition is rare in DIII-D plasmas, likely corresponding to grazing incidences with the resonance location or a breakdown in the Gaussian beam approximation. There is also a large scatter for small values of $A_{ECH}$, suggesting TorbeamNN struggles more at predicting absorption in low peak absorptions. This could correspond to large $\rho$ locations, where the peak ECH power density will be smaller due to larger volume, larger widths which must correspond to shorter peaks, or poor absorption regimes where the total heating is not fully absorbed. While on the surface the $R^2$ values may seem subpar, if we estimate what our real errors will be by looking at the mean average error we get values of: $0.027$ for $\mu_{ECH}$, $0.004$ for $\sigma_{ECH}$, and \SI{0.09}{MW/m^3} for $A_{ECH}$ (per \SI{1}{MW} injected ECH). All of these errors are well within control errors we might expect to see in a real-time PCS and are qualitatively lower than what can be achieved by typical feedforward control. Therefore, these accuracy metrics were deemed acceptable to continue with the ECHO development. 

\subsection{Genetic Algorithm}\label{subsec:genetic_algo}

As explained in \autoref{fig:echo-overview}, a genetic optimizer is used to find the mirror angle and duty cycle for each gyrotron to best match the target deposition profile. This is an optimization algorithm inspired by natural selection, where a population becomes stronger each generation through combination, random mutations and survival of the fittest individuals. By analogy, we define an `individual' to be the collection of all $N$ gyrotron mirror angles and duty cycles. 

While the full algorithm pseudocode is provided in \appautoref{app:pseudocode}, here we provide a written simplified description of the process. A population is first created through randomly selected values from the pre-computed TorbeamNN lookup table, paying attention to enforce the gyrotron ordering. This population will then be altered for multiple generations until an optimized best individual is found. 

For each generation, the cost of each individual in the population is calculated. The cost is defined as the mean-square-error between the target ECH profile and the profile that this individual produces. This is the metric used to assess the quality of optimization. An ``elite percentage'' of the best individuals is then saved, while the rest of the population is filled up with the following steps inspired by evolution. 

To encourage diversity, which is advantageous for finding better local minima, the rest of the population is first filled by ``parents'', which are chosen through tournament selection (see \autoref{alg:tournament}). Each set of parents produces two children, which are a random linear combination of the parents' angles and duty cycles. Each child's angle is then mutated with a given probability (see \autoref{alg:mutate}), meaning the angle is changed randomly within the bounds of adjacent angles.

These children replace the parents to fill the rest of the population, resulting in a new generation of elite individuals and new children. Once the user-defined number of generations is complete, the best individual is selected and its gyrotron angles and duty cycles are sent to the gyrotron hardware. 

\begin{table*}
    \centering
    \begin{tabular}{c|c|c}
        \textbf{Parameter Name} & \textbf{Description} & \textbf{Typical Value} \\\hline
        \textbf{Population size} & Size of the array of individuals & $250$ \\
        \textbf{Generations} & Number of generations taken to find the best individual & $25$ \\
        \textbf{Mutation rate} & Rate at which children have angles and powers mutated & $0.25$ \\
        \textbf{Tournament size} & Size of each tournament to select best parents  & $10$ \\
        \textbf{Elite percentage} & Percentage of the population that is saved each generation & $0.3$ \\
        \textbf{Inertia} & Percentage of the population that is saved each cycle & $0.5$ \\
    \end{tabular}
    \caption{Input parameters to the genetic optimizer, including a brief description and typical values used in DIII-D.}
    \label{tab:genetic-params}
\end{table*}

For the next real-time cycle (typically the next \SI{50}{ms}), a user defined percentage of the best previous population is copied into the new random population. This ensures that if the target and plasma conditions have not significantly changed, the optimization has 'inertia' and will not change its gyrotron request cycle by cycle unnecessarily. It also still enables optimization for a significant target change through the remaining randomly generated individuals. 

The key parameters used in this algorithm are given in \autoref{tab:genetic-params}. Of note, population size and number of generations each increase the total optimizer time approximately linearly, improving convergence to a local minimum. The other parameters have negligible cost on computation times, but are used to balance tradeoffs in diversity of solutions, encourage exploration of solutions, and to avoid gyrotrons from moving unnecessarily every timestep. 

\subsection{PCS Implementation}\label{subsec:PCS}

TorbeamNN was converted from \texttt{Python} to \texttt{C} using \texttt{keras2c}\cite{conlin_keras2c_2021} to make it compatible with the DIII-D PCS. A single TorbeamNN run took $\approx$\SI{63}{\micro s} using the DIII-D PCS hardware. To further increase speed, these computations were parallelized with a PCS compatible multi-threading library\cite{rothsteinParallelizedRealtimePhysics2026}. With this speed-up, an example timestep was able to run $245$ TorbeamNN calculations across $50$ CPU threads in $\approx$\SI{0.32}{ms}. For reference, running the optimized real-time \texttt{TORBEAM} for the same amount of runs across the same cores would take $\approx$\SI{50}{ms}, a speed-up of more than $100$ times. 

ECHO was then connected to \texttt{PACMAN}\cite{rothstein_enabling_2025}, which provides the time-dependent real-time target ECH deposition profiles. While the target may be a feedforward trajectory, \texttt{PACMAN} contains controllers with more advanced applications, such as mode surface tracking or an impurity pumpout algorithm. 

The genetic optimizer was implemented as described in the pseudo-code of \appautoref{app:pseudocode}, and used parameters given in \autoref{tab:genetic-params}. These are empirically determined values which achieved high accuracy, rapid response to changes and fast computation of $\approx$\SI{17}{ms} per cycle, making the genetic optimizer the bottleneck for ECHO speed. Further tuning of the genetic algorithm parameters could have reduced this time, but overall timing and performance was deemed acceptable and not further optimized. The timing of the genetic algorithm could be reduced by incorporating multi-threading into certain parts of the optimization, but the overall algorithm is fairly sequential and cannot be fully parallelized. While not needed for our applications done here, to achieve a significant speed-up other optimization algorithms could be explored. 

Overall, a CPU cycle time of \SI{50}{ms} was used in this experiment. This was decided considering the plasma equilibration timescales and the mirror movement speed is slower or of the order of \SI{50}{ms}. Later experimental applications of ECHO used a \SI{20}{ms} cycle time.

\subsection{Assumptions and Simplifications}\label{subsec:assumptions}

The ECHO algorithm contains several simplifications applied for computational speed and code simplicity. A full ECH control problem description with minimal simplifications is provided in \appautoref{app:full-ECH-problem}. A key assumption in ECHO is that gyrotron powers and angles can be changed instantaneously. While gyrotron power can be rapidly changed through modulation, mirrors can take $\approx$100-\SI{200}{ms} if a large angle change is requested. This means there may be a temporary offset between ECHO commands and truly achieved ECH profiles if the targets change significantly in a single timestep. The impact of the mirror steering time is minimized by enforced gyrotron ordering which avoids large mirror movements.

Another simplification made for real-time use is the discretization of angles and the pre-computation of the TorbeamNN lookup table. While a better solution may be found by allowing the genetic algorithm to guess random continuous angles and run TorbeamNN for each selection, the chosen simplification significantly speeds up the optimization. The discrete angles, typically using step sizes of $0.25^\circ$, are small enough to be negligible compared to the errors involved in ray tracing, including profile and equilibrium reconstruction. 

Finally, the optimization algorithm is constrained by the requirement of a fixed gyrotron order. This means the angle of Gyrotron 1 is always less than Gyrotron 2, and so on. This is chosen to avoid an issue where the optimizer may flip the order of gyrotrons each timestep while still producing the same total profile. By constraining the order, the gyrotrons move only the required amount, reducing the burden on hardware and impact on the plasma. This has the added benefit for the user to force more powerful or weaker gyrotrons to be closer to the plasma core or plasma edge as desired. A disadvantage is that this slows down exploration for the genetic algorithm, as the gyrotrons stuck in between others have limited ability to adjust their angles. However, the experimental results presented in \autoref{sec:applications} show that ECHO quickly explores new configurations without issue.

\section{Experimental results on DIII-D}\label{sec:applications}

The ECHO algorithm was run in several DIII-D experiments, demonstrating the capabilities of the ECH controller and enabling advanced control-based physics experiments. We begin with an experimental demonstration of ECHO responding to changing target deposition profiles in \autoref{sec:changing_profiles}. We then show a measurement validation of the ECHO model in real-time by comparing to electron cyclotron emission (ECE) measurements in \autoref{sec:ECE}. Next, ECHO is applied to discharges with changing plasma conditions, showing the algorithm's real-time adaptability in \autoref{sec:dynamic_conditions}. Finally in \autoref{sec:fault_handling}, we show ECHO effectively responding to gyrotron faults by redistributing other gyrotrons. 

\subsection{Basic demonstration: changing ECH deposition profiles}\label{sec:changing_profiles}

\begin{figure*}
    \centering
    \includegraphics[width=0.75\textwidth]{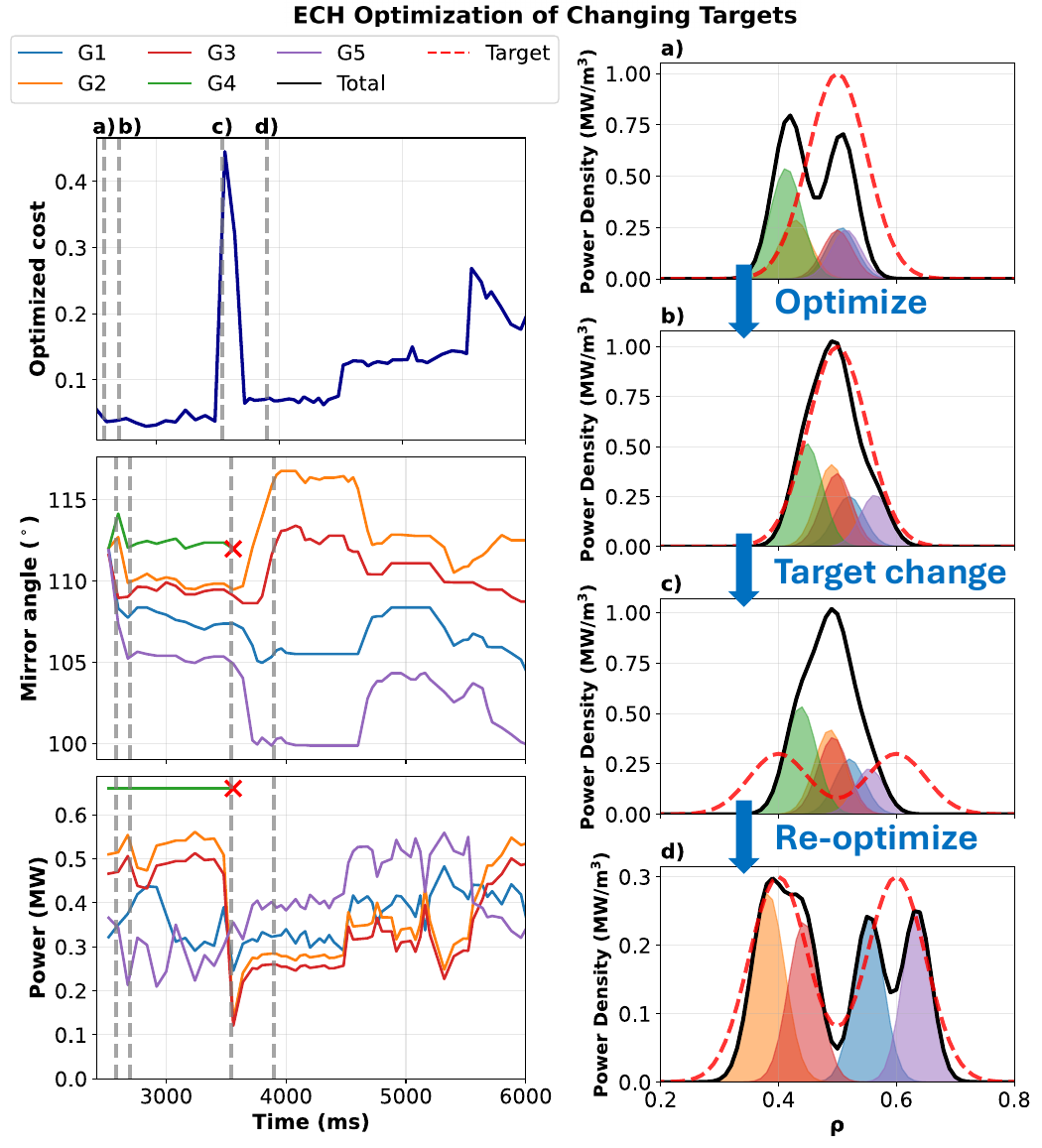}
    \caption{DIII-D shot 205838 demonstrates the flexibility of ECHO, achieving multiple targets rapidly by controlling five gyrotrons. In plots on the right, the targets (red dashed line) are compared to individual gyrotron deposition (colored Gaussians), and the total ECH deposition (black line). At \SI{3600}{ms}, gyrotron 4 fails but the optimizer uses remaining gyrotrons to handle the fault. The optimized cost represents how well the ECH deposition matches the target.}
    \label{fig:changing_profiles}
\end{figure*}

A basic demonstration of the ECHO capabilities used in a DIII-D discharge is shown in \autoref{fig:changing_profiles}, where several ECH deposition profile targets are achieved in a single discharge. The figure shows that the profiles produced by the ECHO algorithm quickly converge to the requested targets. The variety of targets requested here probes both the speed and the flexibility of the algorithm. Control begins at \SI{2580}{ms}, where the gyrotrons are distributed in an unoptimized pattern shown at time a). At \SI{2700}{ms}, all five gyrotrons are quickly redistributed to closely match the target, shown at time b). The target is then changed to a more complex two-peaked distribution, while at the same time gyrotron 4 is lost coincidentally. The optimized cost rises as ECHO searches for a new solution without a gyrotron as seen at time c), but quickly reaches an optimized distribution at time d). 

\autoref{fig:changing_profiles} also shows the optimized cost, measuring the mean-square error between the ECHO optimal profiles and the input target, as well as the angles and powers of each gyrotron. The angles and powers shown are the measured values rather than the values commanded by ECHO. This is why the angles can take $\approx$\SI{150}{ms} to reach a steady value, but powers can be changed much faster. The optimized cost after timestamp \textbf{c)} shows that ECHO quickly finds the mirror angles and profiles, but the true ECH profiles take longer to move large distances. 

This demonstration shows the applicability of ECHO for rapidly changing gyrotron tasks mid-shot, such as shaping the current profiles or to suppress tearing modes, where the total ECH deposition profile will change greatly. When the ECH deposition profile target suddenly changes, ECHO reacts quickly to adjust the available gyrotrons to achieve the target quickly and robustly. 

\subsection{Experimental validation through ECE measurements}\label{sec:ECE}

To validate the real-time control performance achieved in these experiments, \texttt{TORBEAM} and TorbeamNN calculations are compared to deposition profiles extracted from ECE measurements via four inverse methods recently cross-validated for DIII-D by Slief \textit{et al.}\cite{slief_quantifying_2023}.  The profiles obtained from these methods agree with the deposition locations predicted by \texttt{TORBEAM} while capturing the physical beam broadening, as shown in \autoref{fig:4methods}.  

While all four inverse methods provide valuable insights, each presents distinct trade-offs. The Break-in-Slope (BIS) method provides a fast, model-independent time-domain baseline, but it is vulnerable to noise and often overestimates beam width due to diagnostic smearing. Maximum Likelihood Estimation (MLE) effectively handles noisy data by using a robust statistical framework, but its accuracy relies heavily on a specific thermal transport model. Frequency Domain Least Squares (FDLS) excels at isolating the heating response from background noise, though it strictly requires a periodic source. 

Thus, this section focuses on results from the Flux Fit (FF) method due to its consistency and physically rigorous approach while the remaining methods are presented in \appautoref{app:ECE}. The FF method is a self-consistent numerical approach designed to calculate perturbative effects from a modulated heating source in this case the gyrotrons. It simultaneously fits the ECH power deposition profile $\tilde{q}_{\text{ECH}}$ and the underlying heat transport coefficients (diffusive $\chi_e$ and convective $\mathbf{V}$) by solving the linearized heat transport equation:
\begin{equation}
    \frac{3}{2} n_e \frac{\partial \tilde{T}_e}{\partial t} = \nabla \cdot \left[ n_e \left( \chi_e \nabla \tilde{T}_e - \mathbf{V} \tilde{T}_e \right) \right] + \tilde{q}_{\text{ECH}}
\end{equation}
where $\tilde{T}_e$ is the perturbed electron temperature and $n_e$ the electron density, respectively. By incorporating the transport dynamics with the deposition estimation, the FF method accounts for the redistribution of heat that occurs during the modulation cycles. 

\begin{figure*}
    \centering
    \includegraphics[width=\linewidth]{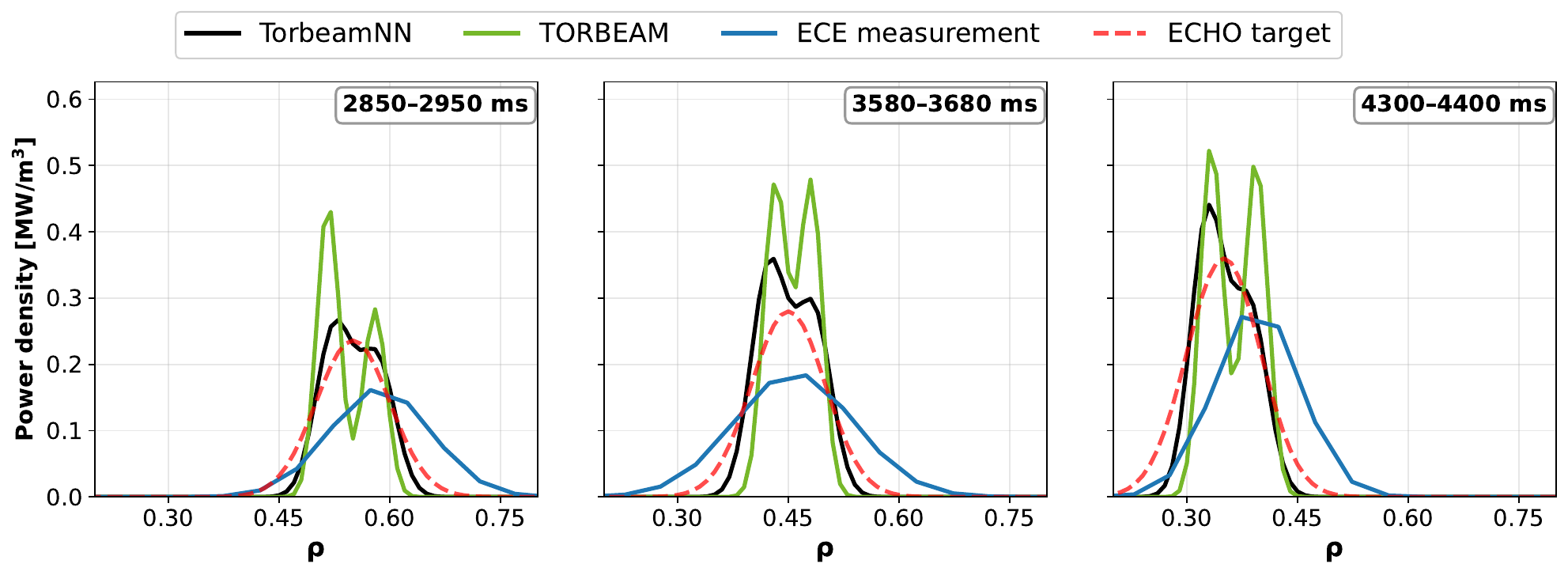}
    \caption{DIII-D shot 205902 presents the power deposition profiles extracted from ECE compared to target and \texttt{TORBEAM} profiles in three time windows. The centers of the peak locations show good agreement. FF extraction reveals additional transport broadening effects, smearing the deposited power into a single, diffuse distribution with a lower peak power density.}
    \label{fig:ECE}
\end{figure*}

For the ECHO validation against ECE data, the gyrotrons were modulated in sync at \SI{246}{Hz}, which enables the deposited power to be extracted from the background electron temperature profile. \autoref{fig:ECE} shows the power deposition profiles extracted from ECE compared to target, TorbeamNN and \texttt{TORBEAM} profiles in three time windows of shot 205902. The centers of the measured ECH profiles are matching or slightly outwards ($\lesssim 0.05$) compared to \texttt{TORBEAM} and the target profiles. This shows that the ECH power is indeed being deposited where ECHO intends over the course of the discharge. Meanwhile, FF profiles are around twice as broad as the ray trace calculated profiles. At the modulation frequencies of this experiment, non-linear transport effects are responsible for this broadening\cite{petty_dpp_2019}. Note that this is no fault of the \texttt{TORBEAM} code, just a limitation of ray tracing calculations to depict realistic broadening.

\subsection{Robustness to changing plasma conditions}\label{sec:dynamic_conditions}

An important use case for real-time ray tracing and optimization is for changing plasma conditions, such as shape changes during ramp-up, or unexpected modes causing significant changes to the plasma profiles. In this section we demonstrate several examples of maintaining a constant deposition profile versus $\rho$ despite a changing plasma.

\begin{figure*}
    \centering
    \includegraphics[width=\linewidth]{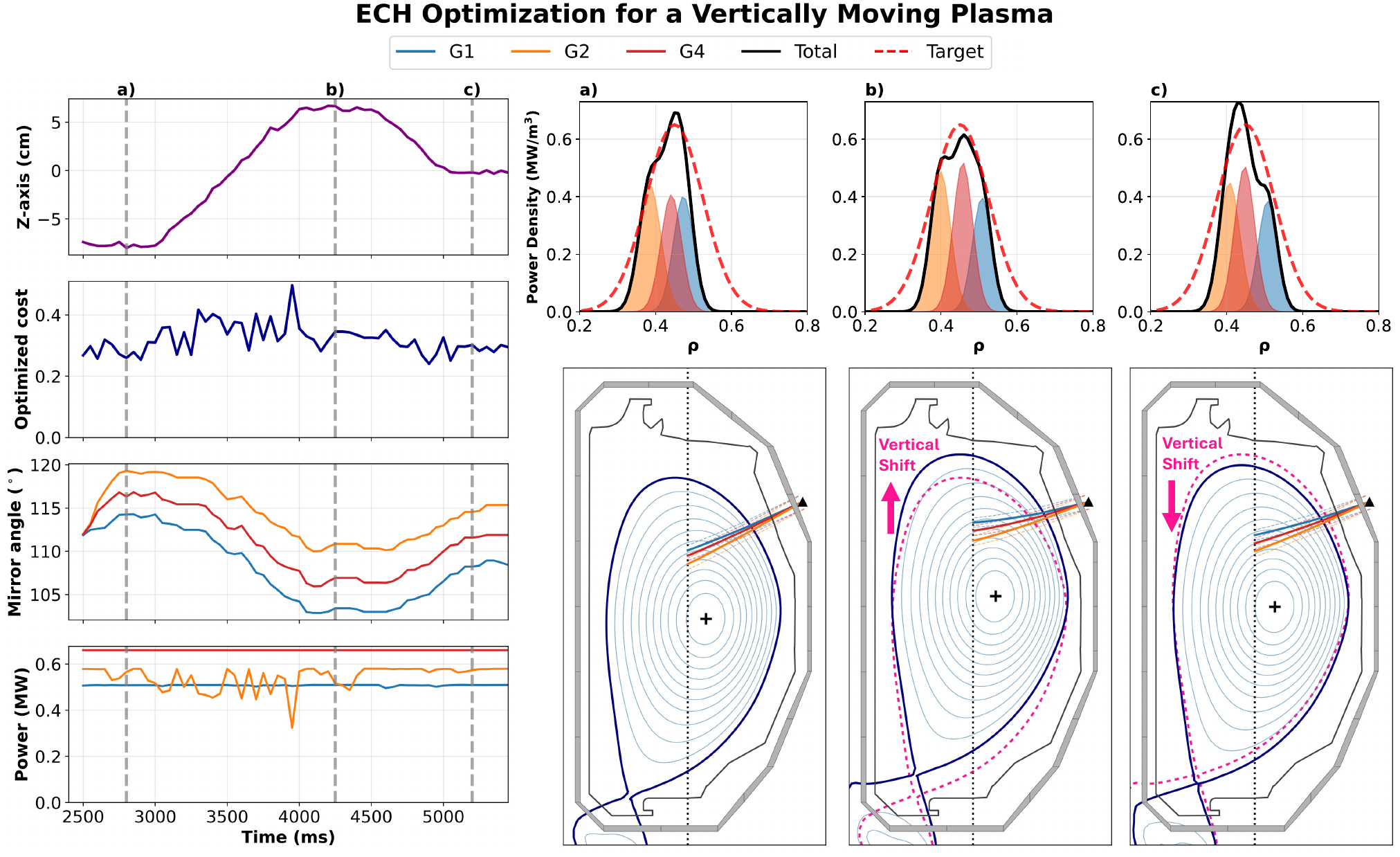}
    \caption{DIII-D shot 205907 demonstrates ECH Optimization for a vertically moving plasma with a constant deposition profile target. The plasma cross-section is significantly displaced upwards from timestep a) to b), and then displaced downwards from b) to c). The physical plasma displacement is visualized in the cross sections in the bottom right, where at each timestep the previous last closed flux surface is shown with a pink dashed line. The ECH resonance location is given as a black vertical dashed line. As the vertical position changes, we see the mirror angle change significantly to track the plasma, and the ECH deposition profiles are successfully kept near-constant. This is also reflected in the Optimized cost timetrace, where the optimization does not lose quality with movement of the plasma.}
    \label{fig:vertical_motion}
\end{figure*}

In \autoref{fig:vertical_motion} we see the ECH deposition profiles at different times, with the plasma moving up and down by over \SI{10}{cm}. As the plasma moves, the mirror angles are adjusted by ECHO to maintain a nearly identical ECH deposition profile with respect to $\rho$. In this use case, the target ECH deposition profile is constant and ECHO correctly adjusts all the gyrotron mirrors and powers to ensure the target is tracked correctly. The powers in this case remain largely unchanged for all gyrotrons because the target profile is unchanged, while the mirror angles must adjust by a large range due to the vertical movement of the plasma. 

\begin{figure*}
    \centering
    \includegraphics[width=0.75\textwidth]{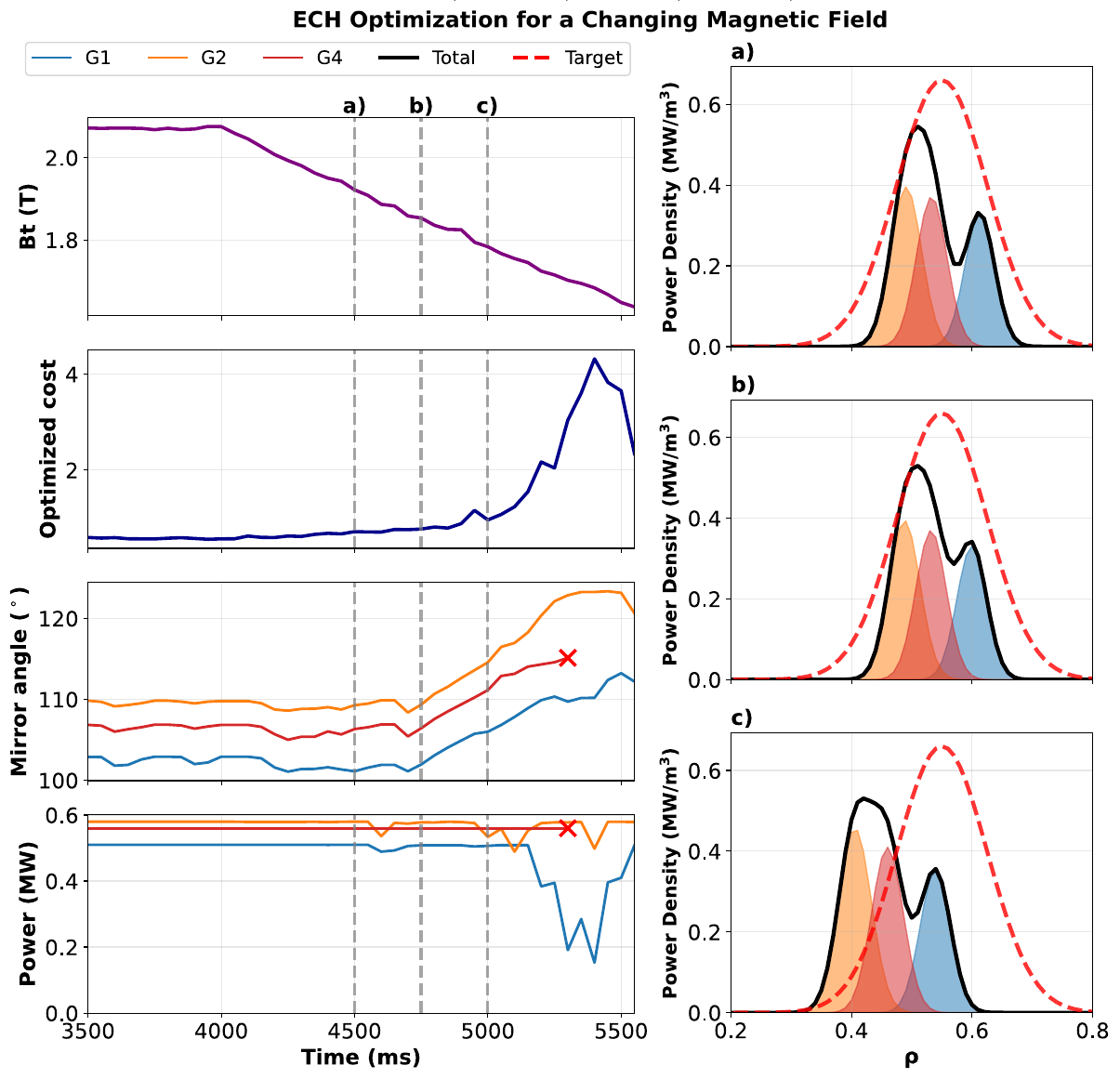}
    \caption{DIII-D shot 205905 where changes to $B_T$ lead to ECHO changing gyrotron locations. The target ECH deposition profile (dashed red line) is slightly larger than the achievable ECH deposition (solid black line). Once $B_T$ decreases significantly, the ECH resonance location has changed to the point that half of the Gaussian target is impossible to reach. This can be seen in timeslice c) and in the increasing optimized cost. Overall, ECHO correctly steers the mirrors to mitigate the $B_T$ changes until it is no longer physically feasible.}
    \label{fig:bt_scan}
\end{figure*}

Another example of changed conditions that affect the ECH deposition is a changing magnetic field ($B_T$) in \autoref{fig:bt_scan}. While this is not a realistic plasma scenario, it serves as a demonstration of ECHO's automatic response to extreme changes. As seen in timeslices \textbf{a)} and \textbf{b)}, ECHO correctly adjusts mirror angles and powers to match the target profile. At timeslice \textbf{c)}, the magnetic field decrease changes the ECH resonance location to the extent that precise aiming is no longer possible. Even in this situation, ECHO continues to provide the best possible match to the target ECH deposition profile. In general, the profile targets in this discharge request higher power density than can be achieved with the three available gyrotrons, so the optimizer can only request the maximal power output for each while aiming accurately. 

These stress tests of rapidly changing conditions demonstrate ECHO's robustness to operating scenario. ECHO can be used in the highly transient ramp-up and ramp-down phases, so long as there is a valid real-time equilibrium and electron temperature and density measurements.

\subsection{Mitigating gyrotron failure}\label{sec:fault_handling}

Gyrotron hardware in DIII-D and most tokamaks can fail and cause the power to go to zero, with a small potential for recovery later in a shot. Typical scenario design relies on pre-determined gyrotrons turning on at pre-defined times. This means unexpected gyrotron failure often results in shots that do not achieve the desired scenario, have significant MHD, or are otherwise not informative for physics experiments. In an FPP, this could equate to a significant loss in fusion gain or large MHD activity leading to an uncontrolled disruption. Accordingly, advanced controllers must be designed to handle actuator failures.  

\begin{figure*}
    \centering
    \includegraphics[width=0.9\linewidth]{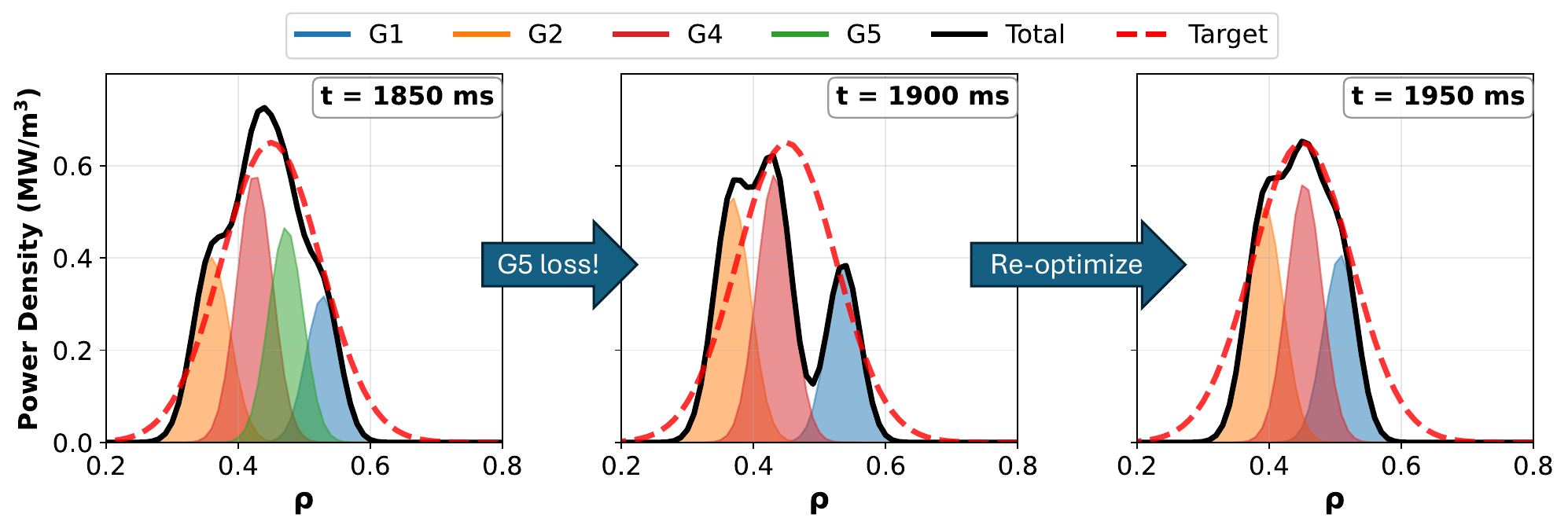}
    \caption{Fault handling demonstration in DIII-D shot 205907. When gyrotron 5 (green) is lost, it opens a gap in the ECH deposition profile. ECHO then re-optimizes the remaining gyrotrons and quickly regains a profile with low error. While this shows real-time ECHO optimization and real requested commands, the mirror hardware did not respond due to an error, meaning this is a simulated demonstration of fault handling.}
    \label{fig:fault-overview}
\end{figure*}

Our ECH control scheme avoids this problem entirely by making the control target, the ECH deposition profile, actuator-independent where ECHO utilizes whatever gyrotron hardware is presently available. In the event of mid-shot actuator failure, ECHO receives an ECH fault signal in real-time and assumes the faulted gyrotron contributes zero power. In the next cycle, ECHO then re-optimizes the remaining functioning gyrotrons to best achieve the target ECH deposition profile in spite of the changes to the available hardware. 

This situation occurred in DIII-D shot 205907 and is shown in \autoref{fig:fault-overview}. In this shot, gyrotron 5 is lost mid-shot causing brief poor performance of the controller. After the ECH fault signal reaches ECHO, the optimizer reacts by changing the angles and powers of gyrotrons 1, 2, and 4. This quick reaction to actuator failure is critical for FPP operation where small lapses in real-time control can lead to large changes in the plasma, leading to poor performance or unstable regimes. While ECHO requested fault-mitigating ECH mirror angles and powers, the mirror hardware did not respond before \SI{2}{s} due to an error. This is therefore a ``simulated'' shot in a sense, demonstrating the capabilities of ECHO, but not reflective of the experimentally produced profiles. An example of fault handling where mirror hardware did respond is shown in \autoref{fig:changing_profiles}, where the G4 failure is well mitigated when re-optimizing. 

ECHO's freedom from hardware dependence comes at the cost of needing to provide an ECH deposition profile target that is reasonably achievable. As seen in \autoref{fig:bt_scan}, the target is already too tall to be achieved and losing a gyrotron in this situation would make the ECH deposition profile extremely poor. While gaining this actuator failure robustness does require some planning on the operator or designing other controllers that provide an ECH deposition target, it provides a reactor-critical layer of robustness that feedforward gyrotron programming cannot achieve.

\section{Discussion and Conclusions}\label{sec:conclusion}

The advanced ECH control framework ECHO has been introduced and demonstrated robustly in DIII-D experiment. Using an ML-based surrogate model of the \texttt{TORBEAM} code, ECHO directly controls the ECH deposition profile and is able to accurately adjust gyrotrons in the presence of dynamic plasma conditions, such as changing $B_T$ or plasma shape, and remains robust when gyrotrons fail unpredictably. ECHO has also been validated against \texttt{TORBEAM} directly as well as ECE measurements to verify the real ECH deposition profiles match the values predicted by the ML surrogate model. 

Future scheduled experiments will apply ECHO to achieve integrated physics and control goals on DIII-D. First, ECHO has been implemented to support sustained RMP ELM suppression through pedestal ECCD and core sawtooth control. Another experiment will demonstrate impurity pumpout through ECH-induced density flattening and impurity shielding through an ECH-sustained tearing mode. These experiments will leverage ECHO to achieve multi-objective control, displaying the full capabilities of the algorithm. Previously, these experimental goals would have required the design of complex controllers for each experiment, but ECHO simplifies the process by automatically organizing the gyrotrons for any ECH goal. This flexibility enables a large variety of future experiments where flexible, failure robust ECH control is required. 

Several upgrades are under development to improve the flexibility of the ECHO and TorbeamNN algorithms. First, the ECCD profile can be predicted alongside ECH. This may require more careful training of the surrogate model, as ECCD is a more complex physics process than ECH. TorbeamNN could also be improved by expanding the Gaussian output representation to include higher moments like skewness and kurtosis. While \texttt{TORBEAM} outputs are well described by a Gaussian, these higher moments would cover edge cases more accurately. 

The ECHO approach to ECH control is well suited for FPPs where failure robustness is critical, and where significantly more gyrotrons will be used. Additionally, all the training for TorbeamNN can be made machine-specific before a tokamak begins operation, so long as a reasonable range of equilibria, $n_e$, and $T_e$ profiles can be assumed and will be available in real-time. Since only offline \texttt{TORBEAM} is required to train the surrogate model and ECH physics does not fundamentally change when moving to an FPP, a well-trained surrogate model can be trusted to predict ECH and ECCD deposition correctly within the parameter space of the training data. Good optimizer parameters could also be estimated offline to the level that minimal tuning was required in experiment, where the main optimizer parameter tuning was to increase computational speed after initial conservative estimates. 

Finally, ECHO can be coupled with offline integrated modeling tools to provide a real-time counterpart to a time-varying ECH deposition profile target. Most integrated modeling suites such as RAPTOR\cite{feliciNonlinearModelbasedOptimization2012b}, TORAX\cite{citrin_torax_2024}, and FUSE\cite{meneghiniFUSEFusionSynthesisa} use the total ECH and ECCD deposition profiles for plasma profile modeling. However, there was no corresponding tool to bring optimized ECH and ECCD deposition profiles to the real hardware, and approximate feedforward aiming and power control was the standard. Using ECHO, offline integrated modeling tools can optimize scenarios using ECH profiles that the hardware is capable of, and provides optimized control of the gyrotron trajectories. ECHO closes the gap from offline simulation and optimizer tools to real-time control while also providing essential robustness to actuator failure.

\section*{CRediT Authorship Contribution Statement}

\noindent 
\textbf{H.J. Farre-Kaga}: Lead Author, Genetic Algorithm Lead, Visualization Lead, Experimental Support, Writing - Original Draft. 
\textbf{A. Rothstein}: Lead Author, Experimental Lead, TorbeamNN Lead, Writing - Original Draft.
\textbf{K. Yasoda}: Validation Lead, TorbeamNN Support, Visualization Support, Writing - Original Draft. 
\textbf{J. Lestz}: Experimental Support, Validation Support, Writing - Review \& Editing.
\textbf{N. Chen}: TorbeamNN Support, Writing - Review \& Editing.
\textbf{S.K. Kim}: Experimental Support, Supervision, Project Administration.
\textbf{A. Jalalvand}: Supervision, Project Administration, Funding Acquisition.
\textbf{E. Kolemen}: Supervision, Project Administration, Funding Acquisition.

\section*{Acknowledgments}

This material is based upon work supported by the U.S. Department of Energy, Office of Science, Office of Fusion Energy Sciences, using the DIII-D National Fusion Facility, a DOE Office of Science user facility, under Award DE-FC02-04ER54698. Additionally, this material is supported by the National Science Foundation Graduate Research Fellowship under Grant No. DGE-2039656 and by the U.S. Department of Energy, under Awards DE-SC0015480 and DE-AC02-09CH11466. 

\section*{Disclaimer}

This report was prepared as an account of work sponsored by an agency of the United States Government. Neither the United States Government nor any agency thereof, nor any of their employees, makes any warranty, express or implied, or assumes any legal liability or responsibility for the accuracy, completeness, or usefulness of any information, apparatus, product, or process disclosed, or represents that its use would not infringe privately owned rights. Reference herein to any specific commercial product, process, or service by trade name, trademark, manufacturer, or otherwise does not necessarily constitute or imply its endorsement, recommendation, or favoring by the United States Government or any agency thereof. The views and opinions of authors expressed herein do not necessarily state or reflect those of the United States Government or any agency thereof.

\section*{Data availability}
All relevant DIII-D data supporting the findings of this study are
available from the DIII-D National Fusion Facility, which is operated
by General Atomics for the U.S. Department of Energy. Access to DIII-D
data requires following the user protocols described on the DIII-D
website at \url{https://d3dfusion.org/become-a-user/}.

\section*{Code availability}
The source code for the models developed in this work is publicly available at \url{https://github.com/PlasmaControl/ECH-OPT}.

\clearpage
\appendix
\section{ECH Control Problem Formulation}\label{app:full-ECH-problem}

\begin{table*}
    \centering
    \begin{tabular}{c|c}
        \textbf{Variable} & \textbf{Description} \\\hline\hline
        $t$ & Discretized time index where $T$ is some fixed time horizon. \\\hline
        $\rho$ & Spatial location in the plasma cross-section \\\hline
        $\theta_i(t)$ & Gyrotron mirror angle for gyrotron $i$ (control actuator) trajectory in time. \\\hline
        $H_i$ & Maximum power of gyrotron $i$ given in \SI{}{MW}. \\\hline
        $s_i(t)$ & Power modulation of gyrotron $i$ where $s_i(t)\in[0,1]$. \\\hline
        $P(t)$ & Plasma state at time $t$. If $T$ is small enough, we can assume $P(t)=\textrm{constant}$. \\\hline
        $f(\theta_i(t),P(t),\rho)$ & \thead{ECH/ECCD deposition profile at spatial location $\rho$ when gyrotron $i$ is aimed at angle $\theta_i(t)$ with \\ power modulation $s_i(t)$ given plasma state $P(t)$. This is computed by the ML surrogate model TorbeamNN.} \\\hline
        $Y(t,\rho)$ & Target ECH or ECCD profile at spatial location $\rho$ at a given time $t$.
    \end{tabular}
    \caption{Variable descriptions of all terms used in defining the ECH control problem.}
    \label{tab:ECH-problem}
\end{table*}

The full receding time horizon control problem is defined as follows with all variables explained in \autoref{tab:ECH-problem}. Two caveats to the definitions are the assumption made of the fixed time horizon $T$ and the range of power modulation $s$. We could consider $T=\infty$, however DIII-D plasma discharges typically last $\leq$\SI{6}{s} so considering $T=\infty$ would not be truly accurate. For a truly steady state fusion power plant where plasma shots may last on the order of minutes to hours, $T=\infty$ could be a more reasonable choice. For $s$, it is not exactly the whole range $s_i(t)\in[0,1]$, but we will return to discuss this later in the discussion of constraints when discussing how ECH power modulation is achieved. 

First the control actuator trajectory is defined by
\begin{align}
    u(t)&=\begin{bmatrix}
        \theta_1(t) & s_1(t) \\
        \vdots & \vdots \\
        \theta_{N_{gyro}}(t) & s_{N_{gyro}}(t)
    \end{bmatrix}
\end{align}
where $N_{gyro}$ is the total number of available gyrotrons. The actuator value at a specific time is then used to compute the total ECH deposition profile at a given $\rho$ with
\begin{align}
    F\left(u(t),t,\rho\right)&=\sum_{i=1}^{N_{gyro}}f(\theta_i(t),P(t),\rho)\times H_i\times s_i(t)
\end{align}
where $f(\theta_i(t),P(t),\rho)$ is computed by TorbeamNN. This is used to define the total cost function
\begin{align}
    J(u_i(t), t=T)&=\int_0^T\int_0^1\left(F(u(t),t,\rho)-Y(t,\rho)\right)^2\mathrm d\rho\mathrm dt 
\end{align}
which represents the total cost over spatial dimension $\rho$ from $t=0$ to $t=T$. This defines our minimization problem below with the goal of finding the optimal actuator trajectory $u^\star$.
\begin{align}
    u^\star(t)=[\theta^\star(t),s^\star(t)]&=\argmin_{u_i}J(u_i(t),t=T)
\end{align}

\subsection{Other Cost Function Considerations}
Moving the mirrors themselves is a costly action because in the timescales of tokamak control, ECH mirror steering is quite slow ($\approx$\SI{100}{ms}) when compared to most other real-time control ($\approx$\SI{10}{ms}). A reasonable modification would be to include a term penalizing $\dot\theta_i(t)$ to discourage the controller from large changes in mirror angle. 

An additional consideration is the time-dependence of this optimization. The current cost function would find extremely poor ECH deposition profiles acceptable if $T$ is large and an optimal solution is achieved for a long enough time. The balance in choice of time horizon $T$ is important because it must be long enough to provide a path to an optimal configuration, but not so long as to enable poor performance in earlier times. Poor performance in earlier times could be problematic for plasma stability if the controller is being used for tearing mode (TM) control and poor intermediate error allows a TM to appear and grow. Besides correct selection of $T$, incorporating a term in the cost function to weight earlier times as more important could be used to achieve this goal, say scaling the integrand of $J(u(t),t)$ by a constant times $(T-t)$ or a similarly time-dependent term. 

\subsection{Optimization Degrees of Freedom and Constraints}
The optimization has $2N_{gyro}$ total degrees of freedom since each gyrotron has a first degree of freedom for each ECH mirror angle $\theta_i$ and a second for each gyrotron duty cycle $s_i$. At present, DIII-D has $N_{gyro}=6$ with plans to increase to above $10$ in the next few years, although in experiment only $5$ gyrotrons were operational.

There is only one type of hardware constraint that limits the degrees of freedom and that is the power supplies that power each gyrotron. As described in \autoref{sec:intro}, ECH power modulation is done by turning power supplies on and off at high frequencies, on the order of \SI{1}{ms}. When two gyrotrons are connected to the same power supplies, they must have the same modulations. For example, if gyrotron $1$ and gyrotron $2$ are both connected to the same power supply, they must have $s_1(t)=s_2(t)$ for all $t$. Present DIII-D hardware has $N_{gyro}=6$ and $N_{PS}=4$ power supplies, so two pairs of gyrotrons must be paired on power supplies which reduces the degrees of freedom by $2$. With this constraint, the total degrees of freedom is reduced to $2N_{gyro}-(N_{gyro}-N_{PS})=N_{gyro}+N_{PS}$. Note that the power supply hardware only enables $2$ gyrotrons to be attached to a single power supply restricting $N_{gyro}\leq 2N_{PS}$.

The final constraint comes in the form of limiting the allowable gyrotron duty cycles. First, the user can specify limits on each $s_i(t)$, such as forcing $s_i(t)\in[0.5,1]$ or fully constraining to $s_i(t)=1$, which is sometimes necessary if a gyrotron power supply cannot be modulated. The other real-time complication comes from the physical limitation of gyrotron power supplies. Gyrotrons would ideally be modulated as quickly as possible to best average the power, but the power supplies have a maximum modulation frequency of \SI{200}{Hz} to \SI{1000}{Hz} depending on the power supply. In the case of \SI{200}{Hz}, this means a gyrotron must be on for a minimum of \SI{2.5}{ms} before it can be turned off and vice versa. If we average the duty cycle over \SI{10}{ms}, the only allowable duty cycles are $s_i(t)=0$, $s_i(t)=1$, or $0.25\leq s_i(t)\leq0.75$. This turns out to be quite restrictive, so the best solution is to increase the averaging interval to \SI{40}{ms}, which is still far below transport timescales of $\approx$\SI{100}{ms} and should be acceptable to average over. This widens the allowable duty cycles to $s_i(t)=0$, $s_i(t)=1$, or $0.0625\leq s_i(t)\leq0.9375$. While an optimizer could be designed to request only these allowable values, our genetic optimizer just rounds to the closest allowed duty cycle value and we do not see any significant change in performance. Future gyrotrons should be able to better handle modulation up to \SI{1}{kHz}, which further reduces the impact of this gyrotron modulation issue.

\section{Genetic Algorithm Pseudo-Code}\label{app:pseudocode}
More comprehensive pseudocode for the genetic algorithm is provided to explain how each stage of the optimizer works. In \autoref{alg:echo} the overall flow of the optimizer is given. This high level flow involves initializing the population, then repeatedly selecting parents through the Tournament selection procedure (\autoref{alg:tournament}), then taking the resulting parents and Crossing over (\autoref{alg:crossover}), before finally adding extra randomness by Mutating the result (\autoref{alg:mutate}). The description of these algorithms includes details on any requirements specific to handling the gyrotron duty cycles and mirror angles constraints. Example values used in experiment are given on the previously shown \autoref{tab:genetic-params}.

\begin{algorithm}[H]
    \caption{ECH Optimizer}
    \label{alg:echo}
    \begin{algorithmic}[1]
        \State \textbf{Inputs:} Target profile $Y$, Population size $N_{pop}$, Generations $N_{gen}$, Elite fraction $p_e$, Mutation rate $p_m$, Inertia $I$
        \State \textbf{Output:} Best gyrotron configuration (angles and duty cycles for each gyrotron)
        \Statex 
        \State $N_{elite} \gets \text{integer}(p_e \times N_{pop})$
        \State \textbf{// 1. Initialize Population (with Inertia)}
        \If{first real-time cycle}
            \State $population \gets N_{pop} \text{ random individuals (preserving gyrotron order)}$
        \Else
            \State $N_{kept} \gets \text{integer}(I \times N_{pop})$
            \State $kept\_inds \gets \text{Top } N_{kept} \text{ individuals from previous cycle}$
            \State $new\_inds \gets (N_{pop} - N_{kept}) \text{ new random individuals}$
            \State $population \gets kept\_inds + new\_inds$
        \EndIf
        \For{\textbf{each} $ind \in population$}
            \State $ind.cost \gets \text{MSE}(ind.profile, Y)$
        \EndFor
        \State \textbf{// 2. Main Evolutionary Loop}
        \For{$g = 1 \dots N_{gen}$}
            \State Sort $population$ in ascending order of $cost$
            \State \textbf{// 3. Save the elite individuals}
            \State $next\_population \gets population[1 \dots N_{elite}]$
            \State \textbf{// 4 \& 7. Fill the rest of the generation}
            \While{$\text{length}(next\_population) < N_{pop}$}
                \State $parent_1 \gets \textsc{TournamentSelect}(population)$
                \State $parent_2 \gets \textsc{TournamentSelect}(population)$
                \State \textbf{// 5. Crossover (random linear combinations)}
                \State $child_1, child_2 \gets \textsc{Crossover}(parent_1, parent_2)$
                \State \textbf{// 6. Mutate (randomize within adjacent bounds)}
                \State $\textsc{Mutate}(child_1, p_m)$
                \State $\textsc{Mutate}(child_2, p_m)$
                \State $child_1.cost \gets \text{MSE}(child_1.profile, Y)$
                \State $child_2.cost \gets \text{MSE}(child_2.profile, Y)$
                \State Append $child_1$ to $next\_population$
                \If{$\text{length}(next\_population) < N_{pop}$}
                    \State Append $child_2$ to $next\_population$
                \EndIf
            \EndWhile
            \State $population \gets next\_population$
        \EndFor
        \State Sort $population$ in ascending order of $cost$
        \State \Return $population[1]$ \Comment{Return the best individual}
    \end{algorithmic}
\end{algorithm}

\begin{algorithm}
    \caption{Tournament Selection (\textsc{TournamentSelect})}
    \label{alg:tournament}
    \begin{algorithmic}[1]
        \State \textbf{Inputs:} Population $P$, Tournament size $N_{tour}$
        \State \textbf{Output:} Selected parent individual 
        \Statex
        \State $pool \gets N_{tour} \text{ random individuals from } P \text{ (with replacement)}$
        \State Sort $pool$ in ascending order of $cost$
        \State \Return $pool[1]$ \Comment{Return the individual with the lowest cost}
    \end{algorithmic}
\end{algorithm}

\begin{algorithm}
    \caption{Crossover (\textsc{Crossover})}
    \label{alg:crossover}
    \begin{algorithmic}[1]
        \State \textbf{Inputs:} Parent individuals $P_1$ and $P_2$, Number of gyros $N_{gyros}$, Number of power supplies $N_{ps}$
        \State \textbf{Output:} Child individuals $C_1$ and $C_2$
        \Statex
        \State Initialize empty individuals $C_1, C_2$
        \State \textbf{// Blend Gyrotron Angles}
        \For{$i = 1 \dots N_{gyros}$}
            \State $\alpha \gets \text{Random float in } [0.0, 1.0]$
            \State $C_1.angles[i] \gets \alpha \times P_1.angles[i] + (1.0 - \alpha) \times P_2.angles[i]$
            \State $C_2.angles[i] \gets \alpha \times P_2.angles[i] + (1.0 - \alpha) \times P_1.angles[i]$
        \EndFor
        \State \textbf{// Blend Duty Cycles}
        \For{$j = 1 \dots N_{ps}$}
            \State $\alpha \gets \text{Random float in } [0.0, 1.0]$
            \State $C_1.duty[j] \gets \alpha \times P_1.duty[j] + (1.0 - \alpha) \times P_2.duty[j]$
            \State $C_2.duty[j] \gets \alpha \times P_2.duty[j] + (1.0 - \alpha) \times P_1.duty[j]$
        \EndFor
        \State \Return $C_1, C_2$
    \end{algorithmic}
\end{algorithm}

\begin{algorithm}
    \caption{Mutation (\textsc{Mutate})}
    \label{alg:mutate}
    \begin{algorithmic}[1]
        \State \textbf{Inputs:} Individual $ind$, Mutation rate $p_m$, Number of gyros $N_{gyros}$, Number of power supplies $N_{ps}$
        \State \textbf{Output:} Individual $ind$ (mutated in-place)
        \Statex
        \State \textbf{// Mutate Gyrotron Angles (Preserving Physical Ordering)}
        \For{$i = 1 \dots N_{gyros}$}
            \If{\text{Random float in } $[0.0, 1.0] < p_m$}
                \State $lower\_bound \gets \text{GlobalMinAngle}[i]$
                \If{$i > 1$}
                    \State $lower\_bound \gets \max(lower\_bound, ind.angles[i-1])$
                \EndIf
                \State $upper\_bound \gets \text{GlobalMaxAngle}[i]$
                \If{$i < N_{gyros}$}
                    \State $upper\_bound \gets \min(upper\_bound, ind.angles[i+1])$
                \EndIf
                \State $ind.angles[i] \gets \text{Random float in } [lower\_bound, upper\_bound]$
            \EndIf
        \EndFor
        \State \textbf{// Mutate Duty Cycles}
        \For{$j = 1 \dots N_{ps}$}
            \If{\text{Random float in } $[0.0, 1.0] < p_m$}
                \State $ind.duty[j] \gets \text{Random float in } [\text{MinDuty}[j], \text{MaxDuty}[j]]$
            \EndIf
        \EndFor
        \State \Return $ind$
    \end{algorithmic}
\end{algorithm}

\section{Measuring ECH power deposition with ECE}\label{app:ECE}

As the previous \autoref{sec:ECE} focused on just the FF method, the remaining 3 methods as well as a comparison to the FF method are described and presented here.

The Break-in-Slope method is a time-domain analysis technique that evaluates the plasma's immediate thermal response to a sudden step change in applied ECH power. By calculating the discontinuity in the time derivative of the local electron temperature precisely at the moment of the power step ($t_0$), the local deposited power density $q_{\text{ECH}}$ can be derived:
\begin{equation}
    q_{\text{ECH}}(\rho) = \frac{3}{2} n_e(\rho) \Delta \left( \frac{\partial T_e(\rho, t)}{\partial t} \right)_{t_0}
\end{equation}
where 
\begin{equation}
    \Delta \left( \frac{\partial T_e}{\partial t} \right)_{t_0} = \left. \frac{\partial T_e}{\partial t} \right|_{t_0^+} - \left. \frac{\partial T_e}{\partial t} \right|_{t_0^-}
\end{equation}
and $n_e$ is the local electron density.

Maximum Likelihood Estimation estimates the transport and deposition parameters $\theta$ by maximizing the likelihood of the observed temperature measurements given the model. Assuming Gaussian noise with variance $\sigma_i^2$, the log-likelihood function is given by:
\begin{equation}
    \ell(\theta) = -\frac{1}{2} \sum_{i=1}^{N} \left[ \frac{(T_{e,\text{meas}}(t_i) - T_{e,\text{model}}(t_i, \theta))^2}{\sigma_i^2} + \ln(2\pi\sigma_i^2) \right]
\end{equation}
By maximizing $\ell(\theta)$, this method effectively handles measurement uncertainties and background noise in ECE measurements.

Frequency Domain Least Squares method transforms the time-series data into the Fourier domain and estimates the power deposition profile by minimizing the least-squares error between the measured frequency-domain temperature $\tilde{T}_{e,\text{meas}}$ and a parameterized transfer function $H$:
\begin{equation}
    J_{\text{FDLS}}(\theta) = \sum_{k} \left| \tilde{T}_{e,\text{meas}}(\omega_k, \rho) - H(\omega_k, \rho, \theta) \tilde{P}_{\text{ECH}}(\omega_k) \right|^2
\end{equation}
This frequency-domain approach helps isolate the fast perturbative ECH response from slower background transport timescales. 

\begin{figure}
    \centering
    \includegraphics[width=\linewidth]{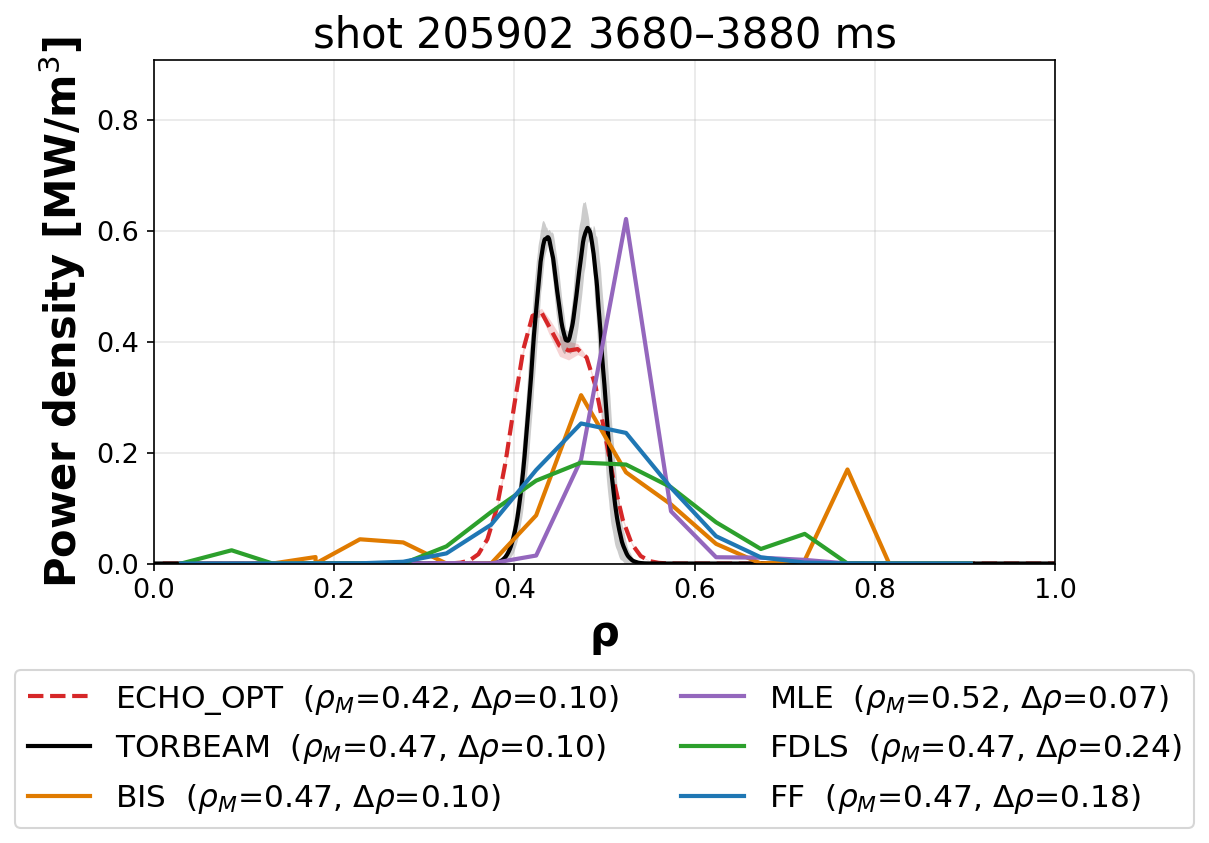}
    \caption{DIII-D shot 205902 power deposition profiles extracted from ECE by 4 different methods compared to target and \texttt{TORBEAM} profiles. Notably, the FDLS and FF extractions reveal substantial physical beam broadening relative to the highly localized \texttt{TORBEAM} prediction. MLE shows an outward radial shift.}
    \label{fig:4methods}
\end{figure}

The four methods are compared in \autoref{fig:4methods} alongside the ECHO target and the \texttt{TORBEAM} offline calculation. All methods show generally similar results, with slightly varying levels of profile broadening. Additionally, the slight outward shift of all measurement methods indicates a high level of perpendicular transport distributing heated electrons to larger values of $\rho$ than where the ray trace codes predict.

\clearpage
\section*{References}
\bibliographystyle{plainnat}
\bibliography{ech_control}

\end{document}